\documentclass[aps,pre,superscriptaddress,twocolumn,twoside,a4paper,floatfix]{revtex4}
\usepackage{hyperref,graphicx,braket, amsmath,amssymb,mathpazo,xfrac,array }

\DeclareMathAlphabet{\mathpzc}{OT1}{pzc}{m}{it}

\newcommand{\ba}{\begin{eqnarray}}
\newcommand{\ea}{\end{eqnarray}}

\begin{document}
\title{The Complete Quantum Cheshire Cat}
\author{Yelena Guryanova}\affiliation{H. H. Wills Physics Laboratory, University of Bristol$\text{,}$ Tyndall Avenue, Bristol, BS8 1TL, United Kingdom}
\author{Nicolas Brunner}\affiliation{H. H. Wills Physics Laboratory, University of Bristol$\text{,}$ Tyndall Avenue, Bristol, BS8 1TL, United Kingdom}
\author{Sandu Popescu} \affiliation{H. H. Wills Physics Laboratory, University of Bristol$\text{,}$ Tyndall Avenue, Bristol, BS8 1TL, United Kingdom}

\begin{abstract}
We show that a physical property can be \emph{entirely} separated from the object it belongs to, hence realizing a complete quantum Cheshire cat.
Our setup makes use of a type of quantum state of particular interest, namely an entangled pre- and post-selected state, in which the pre- and post-selections are entangled with each other. Finally we
propose a scheme for the experimental implementation of these ideas.
\end{abstract}

\maketitle

\section{Introduction}
Quantum mechanics predicts a wide range of astonishing and strikingly counter-intuitive phenomena. Sometimes, it gets even difficult to distinguish
between science and science-fiction, as nicely illustrated for instance by the process of quantum teleportation.

Recently, Aharonov, Popescu and Skrzypczyk \cite{cats} presented another striking situation in which quantum predictions are highly reminiscent of science-fiction literature. Inspired by the adventures of Alice in Wonderland, these authors described a quantum Cheshire cat. In this proposal, a physical property is disembodied from the object it belongs to, much like the 'grin without a cat' in Lewis Carroll's novel. 

In Aharonov et al.'s example, the considered physical system is a polarized photon, sent into an interferometer. Using a judicious arrangement of pre- and post-selection, the physical property, i.e. the polarization, gets separated from the object it belongs to, i.e. the photon. 
While the photon is found with certainty in the left arm of the interferometer, its circular polarization is found only in the right arm of the interferometer.

Importantly, however, in Ref. \cite{cats} the process of 'disembodification' is only partially achieved, in the sense that it is only the circular polarization of the photon that is separated from the photon. The linear polarization is however still 'attached' to the photon, and will hence be detected in the left arm of the interferometer, i.e. the same arm in which the photon is found. This naturally raises the question of whether or not a physical property can be \emph{entirely disembodied} from its physical carrier. Here we answer this question positively, by presenting an example in which a photon is completely separated from all its polarization components, hence representing a complete quantum Cheshire cat. 

The key ingredient in our construction is a quantum state of particular conceptual interest, namely an entangled pre- and post-selected state, in which the pre- and post-selections are entangled with each other \cite{multi1,multi}. Finally we describe an experimental setup, feasible with current technology, for implementing our proposal.

\section{Partial quantum Cheshire cat}

We start by considering the quantum Cheshire cat introduced in Ref. \cite{cats}. A horizontally polarized photon is sent through a Mach-Zehnder interferometer (see Fig.~1). After the first 50:50 beam splitter ($BS_1$), the state of the photon is given by
\ba \label{pre} \ket{\Psi} = \frac{1}{\sqrt{2}}(\ket{L}+\ket{R})\ket{H} \ea
where $\ket{L,R}$ denote the left and right mode of the interferometer, and $\ket{H}$ indicates that the photon is horizontally polarized. This represents the pre-selected state.
The interferometer is equilibrated such that if the state at the second beam splitter ($BS_2$) is $\ket{\Psi}$, then the photon will emerge with certainty in the left output mode. On this same mode we add a polarization measurement device (in the horizontal-vertical basis), consisting of a polarizing beam splitter (PBS) followed by two photon detectors ($D_1$ and $D_2$).

\begin{center}
\begin{figure}[b!]
\includegraphics[scale=0.37]{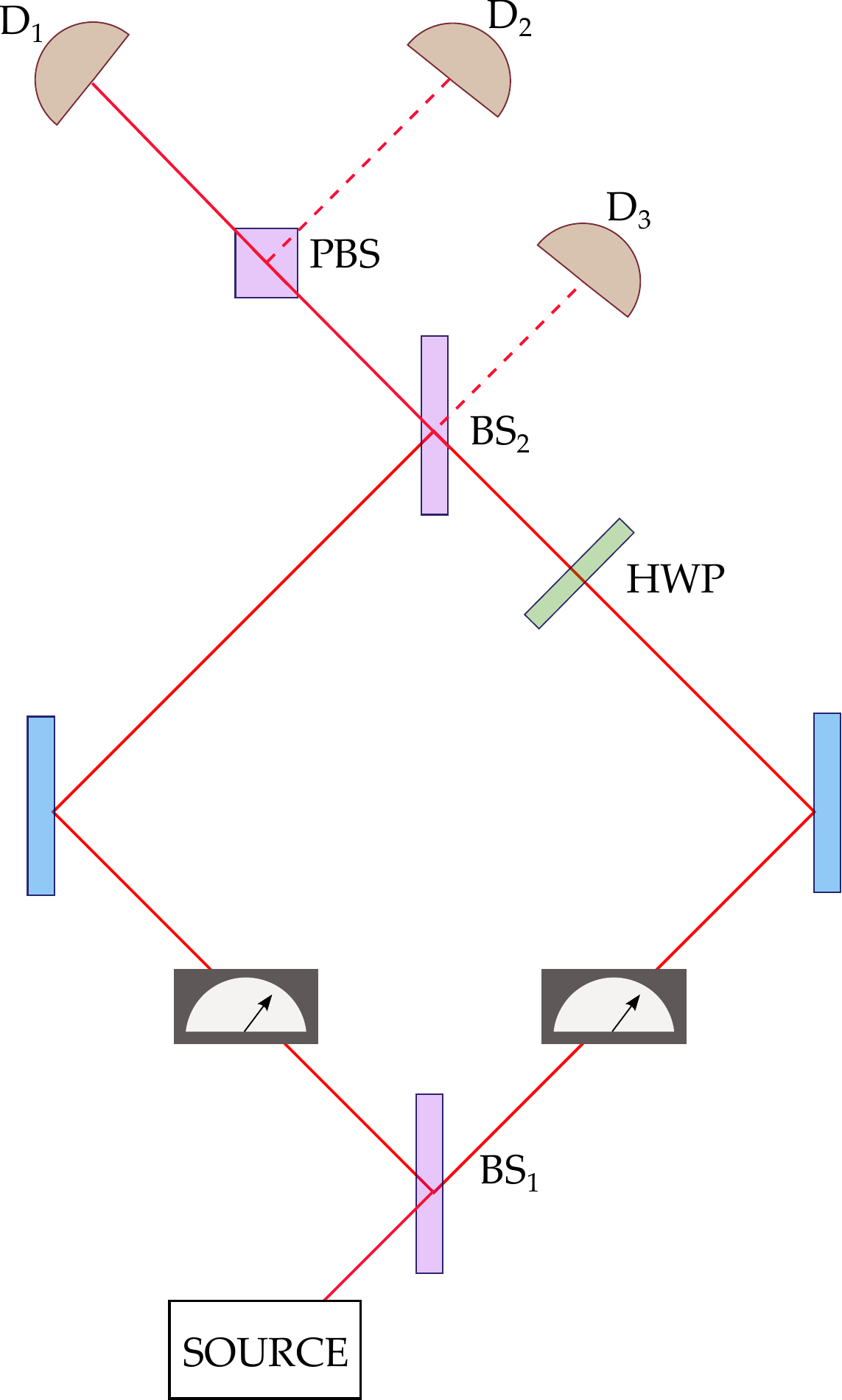}
\caption{Scheme for implementing the partial quantum Cheshire cat \cite{cats}. Measurement devices are placed in the left and right arm of the interferometer, in order to characterize the physical properties of the quantum Cheshire cat, represented by the pre- and post-selected state $\bra{\Phi}\,\ket{\Psi}$.}
\label{aha}
\end{figure}
\end{center}

Next we introduce post-selection. We will consider only those events in which the photon is detected in the horizontal output mode, i.e. the cases in which detector $D_1$ clicks. Here we are interested in the post-selected state inside the interferometer, that is at the location of the measuring devices (see Fig.~1), which is found to be

\ba \label{post} \bra{\Phi} = \frac{1}{\sqrt{2}}(\bra{L}\bra{H} + \bra{R}\bra{V}). \ea
This is our post-selected state, hence the 'bra' notation.
To see this, it is convenient to consider a single photon produced at the location of detector $D_1$ and to evolve it (backwards) to the desired point inside the interferometer. Note the importance of the half-plate (HWP) in the right arm of the interferometer, which flips horizontal and vertical polarization.

Now that we have defined our pre- and post-selected state, i.e. $\bra{\Phi}\, \ket{\Psi}$, we will determine its physical properties by performing weak measurements on it. One may wonder why we do not consider strong (standard) measurements here. This is because strong measurements, which typically disturb the state of the system under measurement, will perturb our Cheshire cat, thus preventing us from observing interesting quantum behaviour \cite{cats,hardyweak}. By using weak measurements, which disturb the system only very little, we can avoid this problem, while still being able to gain information about the system, and thus see interesting quantum phenomena. Note however that in order to achieve non disturbance, we have to pay a penalty: the measurements are imprecise and must hence be repeated many times. The relevant physical value is the average displacement of the pointer.
More precisely, a weak measurement of an observable $A$ on a pre- and post-selected state $\bra{\Phi} \ket{\Psi}$ indicates for $A$ the weak value \cite{AAV}:

\begin{equation}
\langle A\rangle _w = \frac{\bra{\Phi}A\ket{\Psi}}{\braket{\Phi |\Psi}}.
\end{equation}

Here, we consider which-path measurements, represented by the operators
\ba \label{mode}
\ket{L}\bra{L} = \Pi_L  \quad , \quad  \ket{R}\bra{R} = \Pi_R \ea
for the left and right arm respectively. We also consider polarization measurements in each arm, represented by operators of the form

\ba\label{modepol}
\sigma_i^L =\Pi_L \sigma_i = \ket{L}\bra{L} \sigma_i \ea
where the upper index $L$ indicates that the measurement takes place in the left arm, and $\sigma_i$ are the usual Pauli matrices:
\begin{align}
\sigma_z &= \ket{H}\bra{H} - \ket{V}\bra{V} \nonumber \\ 
\sigma_x &=  \ket{H}\bra{V} + \ket{V}\bra{H} \\
\sigma_y &=  -i\ket{H}\bra{V} + i\ket{V}\bra{H} \nonumber
\end{align}
Hence, $\ket{H}$ and $\ket{V}$ are the eigenstates of $\sigma_z$, while the eigenstates of the $\sigma_x$ ($\sigma_y$) are linear diagonal polarizations (circular polarizations).

%In Aharonov's example \cite{cats}, the pre- and post-selected states were 
%\begin{eqnarray}
%\ket{\Psi_1} &=(\ket{L} +\ket{R})\ket{H} \label{ket}\\ 
%\bra{\Phi_1} &= \bra{L}\bra{H}+ \bra{R}\bra{V} \label{bra}
%\end{eqnarray} respectively. (The subscript 1 will become clear later) The following results were obtained:
The weak  values of the observables \eqref{mode} are found to be 
\ba \langle \Pi_L \rangle_w &= \frac{\bra{\Phi} L\rangle \langle L\ket{\Psi}}{\braket{\Phi |\Psi}} = 1\label{res1} \\ 
\langle \Pi_R \rangle_w &= \frac{\bra{\Phi} R\rangle \langle R\ket{\Psi}}{\braket{\Phi |\Psi}} = 0 \label{res2} 
\ea
Thus the photon is found to be in the left arm with certainty.

Now we calculate the weak values of each polarization component in both arms, e.g.
\ba 
\langle \sigma_i^L \rangle_w = \frac{\bra{\Phi} L\rangle \langle L|\sigma_i\ket{\Psi}}{\braket{\Phi |\Psi}} . 
\ea
The results are summarized in Table~1. 
%\begin{align}
%\langle \Pi_L \rangle_w &= \frac{\bra{\Phi_1} L\rangle \langle L\ket{\Psi_1}}{\braket{\Phi_1 |\Psi_1}} = 1\label{res1} \\ 
%\langle \Pi_R \rangle_w &= \frac{\bra{\Phi_1} R\rangle \langle R\ket{\Psi_1}}{\braket{\Phi_1 |\Psi_1}} = 0 \label{res2}\\ 
%\langle \sigma_x^L \rangle_w &= \frac{\bra{\Phi_1} L\rangle \langle L|\sigma_x\ket{\Psi_1}}{\braket{\Phi_1 |\Psi_1}} = 0 \label{res3}\\ 
%\langle \sigma_x^R \rangle_w &= \frac{\bra{\Phi_1} R\rangle \langle R|\sigma_x\ket{\Psi_1}}{\braket{\Phi_1 |\Psi_1}} = 1 \label{res4}
%\end{align}
%which say that the photon is in the left arm, but its x-polarisation is in the right arm! Thus the cat - the photon - is disembodied from the smile - the polarisation.\\

We see that in the left arm, the photon is found with certainty, but both the $x$ and $y$ polarization components vanish.
On the contrary, in the right arm, the $x$ and $y$ polarization components are non-zero, although the photon is never detected.
Thus, the cat (the photon) is disembodied from its grin (the circular and diagonal polarizations, i.e. the $x$ and $y$ components). 

However, from table 1 it is clear that this 'disembodification' process is here only partial. Indeed, the $\sigma_z$ component of the polarization is not vanishing in the left mode, where the photon is found with certainty. Moreover, one may argue that this component corresponds exactly to the initial polarization of the photon, hence making it unclear (at least to the skeptic) whether 'disembodification' really occurs or not.

In the next section, we will remove all doubts by presenting a complete quantum Cheshire cat.

%Here, the polarisation in only one direction was considered for a particular pre-and post-selected state. We will now complete the picture by considering pre- and post-selection and weak measurements of all the polarisation directions and truly eliminating all components of polarisation from the left arm.\\\\

%We can calculate the weak values of all of the spin operators on our pre and post-selection $\ket{\Psi_1}$ and $\bra{\Phi_1} $using the same prescription as in Eqns.(\ref{res1} -\ref{res4}) and using the spin operators defined earlier. The results are shown in Fig. (\ref{tab1}). 

%We see that although there is no $x$ or $y$-component of polarisation in the left arm, there is $z-$polarisation with value 1.  Complete separation will be our challenge in the next section.  

%\begin{figure}[h!]
%\begin{center}
\begin{table}[ht]
\caption{The partial quantum Cheshire cat of Ref. \cite{cats}. The table shows the weak values of polarization measurements in each arm of the interferometer. The operator $\mathbb{I}$ corresponds to a which path measurement. }
\centering
\begin{tabular}
{ |l| >{\centering\arraybackslash}p{0.9cm}| >{\centering\arraybackslash}p{0.9cm}| >{\centering\arraybackslash}p{0.9cm}| >{\centering\arraybackslash}p{0.9cm}| }
\hline 
\multicolumn{5}{|c|}{Partial Cheshire cat  $\quad$ (state $\;\bra{\Phi}$  $\ket{\Psi}$)} \\
\hline
Polarization Meas. &$\sigma_x$& $\sigma_y$ & $\sigma_z$& $\mathbb{I}$   \\  \hline
Left arm & 0&0 &1 & 1 \\ [1ex]\hline
Right arm &1 &i & 0&0  \\ [1ex]\hline
\end{tabular}
\end{table}
%\end{center}
%\label{tab1}
%\end{figure}

\section{Complete quantum Cheshire cat}

Our main ingredient to build a complete Cheshire cat will be a type of quantum states of particular conceptual interest, namely an entangled pre- and post-selected state. 
Such a state consists in a quantum superposition of pre- and post-selected states, for instance of the form
\ba\label{timestate}
 {\bra{\Phi}}\,\ket{\Psi} + {\bra{\Phi'}\,\ket{\Psi'}}
\ea
where $\ket{\Psi}$ and $\ket{\Psi'}$ are pre-selected state (at time $t_1$) and $\ket{\Phi}$ and $\ket{\Phi'}$ are post-selected states (at a later time $t_2$). 

The weak value of observable $A$ on such a state, observed in a measurement performed at an intermediate time $t_1<t<t_2$, is obtained by inserting the corresponding operator in the slots, i.e.

\begin{align}\label{weak}
\langle A \rangle_w = \frac{ \bra{\Phi} A\ket{\Psi} + \bra{\Phi'} A \ket{\Psi'}}{ \braket{\Phi|\Psi} +  \braket{\Phi'|\Psi'}}
\end{align}
Notice that the normalisation of the states in Eqn. (\ref{timestate}) can be omitted, as it does not affect the weak value.

We are now ready to present our complete Cheshire cat. We consider a state of the form \eqref{timestate}, where $\ket{\Psi} $ and $\bra{\Phi}$ are taken as in \eqref{pre} and \eqref{post} respectively, and
\begin{align}
\ket{\Psi'} &= \frac{1}{\sqrt{2}}(\ket{L}+\ket{R}) \ket{V} \label{pre2}\\
\bra{\Phi'} &= \frac{1}{\sqrt{2}} (\bra{L}\bra{V}+\bra{R}\bra{H}) \label{post2}
\end{align} 
Note that the state ${\bra{\Phi'}}\,\,\ket{\Psi'} $ is simply obtained from ${\bra{\Phi}}\,\,\ket{\Psi} $ by flipping the polarization in all terms. In the next section we will discuss how such states can be realized.

Using \eqref{weak}, it is now straightforward to compute the weak values of which path and polarization observables. 
We find again that
\ba
\langle \Pi_L \rangle_w = 1 \quad , \quad \langle \Pi_R \rangle_w = 0 \ea
Thus the photon is found in the left arm with certainty.
Next we determine the polarization component in the left arm 
\ba
\langle \sigma_i^L \rangle_w = 0  \quad \text{for  } i=x,y,z \ea
and in the right arm
\ba \langle \sigma_x^R \rangle_w = 1 \quad , \quad \langle \sigma_y^R \rangle_w = \langle \sigma_z^R \rangle_w =0. \ea
Thus, in the left arm of the interferometer we have a now fully depolarized photon, whereas in the right arm we find a diagonal polarization, i.e along the $x$ axis (see Table~2 for a summary of the results).
The quantum Cheshire cat is now complete, since the photon and the polarization have been completely separated.

\begin{table}[h!]
\caption{The complete quantum Cheshire cat. The photon is fully separated from all its polarization components.}
\centering
\begin{tabular} { |l| >{\centering\arraybackslash}p{0.9cm}| >{\centering\arraybackslash}p{0.9cm}| >{\centering\arraybackslash}p{0.9cm}| >{\centering\arraybackslash}p{0.9cm}| }
\hline 
\multicolumn{5}{|c|}{Complete Cheshire cat $\quad$ (state $\bra{\Phi} \, \ket{\Psi}+\bra{\Phi'} \, \ket{\Psi'}$})  \\
\hline
Polarization Meas. &$\sigma_x$& $\sigma_y$ & $\sigma_z$& $\mathbb{I}$   \\  \hline
Left arm & 0&0 &0 & 1 \\ [1ex]\hline
Right arm &1 &0 & 0&0  \\ [1ex]\hline
\end{tabular}
\end{table}

As a side remark, we note that instead of considering a quantum superposition of pre- and post-selected states, as we did here, one may also consider a mixture of pre- and post-selected states. However, the way in which to prepare such mixture, and more generally the physical meaning of such states is at present not well understood.

\section{Experiment}
We present a simple optical scheme for the implementation of the complete Cheshire cat. We build upon the scheme of Fig.~1, which realized the pre- and post-selected state ${\bra{\Phi}}\,\,\ket{\Psi} $. Clearly, the scheme can be trivially modified to implement the state ${\bra{\Phi'}}\,\,\ket{\Psi'} $, by simply flipping the initial polarization of the photon and post-selecting events where detector $D_2$ clicks (see Fig.~1). But now the question is how can we superimpose these two states, in order to realize the desired entangled pre- and post-selected state of Eq. \eqref{timestate}?
 
In order to do so, we must use an ancilla, which will allow us to 'entangle' the pre- and post-selections. Here this ancilla is an auxillary photon, which is initially entangled with the system, i.e. the Cheshire photon. By recombining the Cheshire photon and the ancilla after the Mach-Zender interferometer, via a joint (entangled) measurement, the pre- and post-selections become entangled hence achieving effectively the desired state.

\begin{figure}[t!]
\includegraphics[scale=0.4]{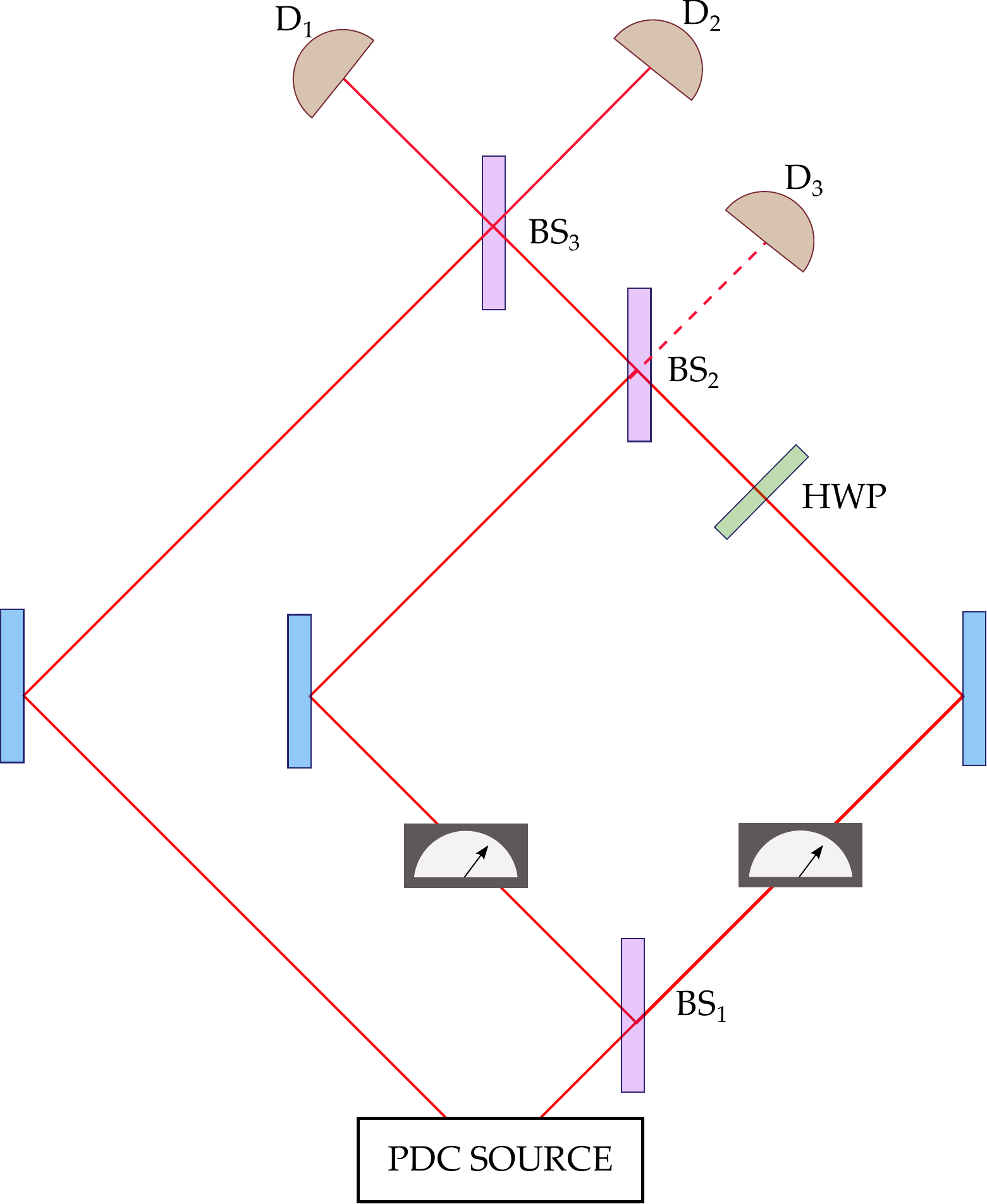}
\caption{Optical implementation of the complete quantum Cheshire cat. The measurement devices placed in the left and right arm of the interferometer allow us to determine the physical properties of the the complete quantum Cheshire cat, given by the entangled pre- and post-selected state $\bra{\Phi}\, \ket{\Psi} + \bra{\Phi^{\prime}}\, \ket{\Psi^{\prime}}$ .}
\label{setup}
\end{figure}

More specifically, let us consider a parametric down-conversion (PDC) source producing pairs of photons entangled in polarization in a singlet state

%To observe the effect described above -- i.e to see the photon in one arm of the interferometer and its polarisation entirely in the other we will need to adjust the experiment described by Aharonov \cite{cats} to included an ancilla. The ancilla allows us to prepare both of the two time states $ \bra{\Phi_1}\quad\ket{\Psi_1} $ and $  \bra{\Phi_2}\quad\ket{\Psi_2}$. \\
%An entanglement source prepares two photons in the entangled state 
\ba \label{singlet}
\ket{\psi_-}=  \frac{1}{\sqrt{2}} ( \ket{H}_S\ket{V}_A - \ket{V}_S\ket{H}_A)
\ea
where the subscripts $S$ and $ A $ denote the system and ancilla respectively. The system, i.e. the Cheshire photon, is sent towards the Mach-Zender interferometer, while the ancilla photon travels along the outer-most arm of the experimental setup (see Fig. (\ref{setup})). 
The MZ interferometer is equilibrated as before. 
Thus the pre-selected state, including the ancilla, is of the form
\ba \label{PRE} \left(\ket{H}_S\ket{L+R} \right)_S\ket{V}_A   -\left(\ket{V}_S\ket{L+R} \right)_S\ket{H}_A \ea

Finally, the system and the ancilla are recombined at $BS_3$. When observing one click in both detectors $D_1$ and $D_2$, the state of the system and ancilla is projected onto the singlet polarization state $\ket{\psi_-}$. 
Hence, the post-selected state, including the ancilla, is of the form
\ba\label{POST} \left(\bra{LH}_S+ \bra{RV}_S \right)\bra{V}_A -\left(\bra{LV}_S+ \bra{RH}_S \right)\bra{H}_A \ea

Since the polarization state of the ancilla is not modified between the initial preparation and final joint measurement, we obtain the pre- and post-selected state inside the MZ interferometer by contracting the ancilla between equations \eqref{PRE} and \eqref{POST}. Hence, we obtain the desired state, i.e. a superposition of pre- and post-selected states of the form 
\ba \bra{\Phi}\,\ket{\Psi} + \bra{\Phi'}\,\ket{\Psi'} \ea

The next important issue is how can one perform experimentally the weak measurements required in order to observe the quantum Cheshire. In other words, one can we measure weakly which path the photon took, and moreover the state of polarization in each mode. 

There are several possibilities which can be explored here. The main idea consists in coupling other degrees of freedom of the photon (the optical beam), to the observables to be measured. For instance, a weak measurement of the path can be implemented by inserting a small piece of glass in (say) the left beam, the effect of which is to slightly displace the beam perpendicularly to its propagation direction (see e.g. \cite{resch,mir}). The profile of the beam then becomes the pointer of our 'which-path' measurement. Then, by replacing the detectors ($D_1$ and $D_2$) by CCD cameras, we can measure the profile of the beam, that is, the displacement of our pointer, hence gaining information about which path the photon took.

Concerning the weak measurements of polarization, they could be implemented in a similar fashion, using for instance a birefringent element (see \cite{cats}), which couples (very weakly) a particular polarization component to a displacement of the beam perpendicular to the direction of propagation. Again, using CCD cameras, this displacement can be monitored. 

Another possibility would consist in coupling a particular polarization component to the temporal mode of the photon. This can be done conveniently using birefringence. In the regime where the birefringence is small compared to the temporal width of the photon's wavefunction, the measurement of polarisation will be weak \cite{ritchie,nb}. Then, information about the polarization can be retrieved by performing a time-of-arrival measurement.

\bigskip

\section{Discussion}
We have described a complete quantum Cheshire cat, thus showing that a physical property can be entirely disembodied from the object it belongs to. In our example, a photon gets completely separated from all its polarization components. We also presented an optical setup for realizing the 'complete quantum Cheshire cat', which should be feasible using current technology.

In our scheme the Cheshire cat is witnessed using weak measurements. Since we use pre- and post-selected states, information is retrieved in weak values of quantum observables. Such weak values have been observed experimentally, and turn out to be particularly useful in the context of precision measurements, where they lead to an amplification effect.

Interestingly, in the context of precision measurements based on weak measurement amplification discussed so far, the importance of the quantum nature of the weak value is however not clear. Indeed, all experiments realized so far can in fact be described in terms of classical optics, as they do not make use of entanglement or any other genuinely quantum aspects of light. Hence, those schemes should in fact be referred to as 'quantum-inspired' metrology. Notably, this is not the case with our complete Cheshire cat. Indeed here, the pre- and post-selections are entangled with each other, which is a genuinely quantum property. Thus, there is no classical analogue to our scheme. It would therefore be interesting to investigate the potential for metrology of our complete quantum Cheshire cat.

\textbf{Acknowledgements} We acknowledge financial support from the UK EPSRC, the EU grant Q-ESSENCE, the Templeton Foundation, and an advanced ERC grant (NLST).

\end{document}